\documentclass[preprint,11pt]{article}

\usepackage{blindtext}
\usepackage{moreverb,url}
\usepackage{float}
\usepackage{physics}
\usepackage{stmaryrd}
\usepackage{booktabs}
\usepackage{rotating}
\usepackage{jmlr2e}
\usepackage{amsmath}
\usepackage{bm}

\usepackage{lastpage}
\jmlrheading{23}{2022}{1-\pageref{LastPage}}{1/21; Revised 5/22}{9/22}{21-0000}{}

\ShortHeadings{}{}
\firstpageno{1}

\begin{document}

\title{Randomized Basket Trial with an Interim Analysis (RaBIt) and Applications in Mental Health}

\author{\name Sahil Swapnesh Patel \email \\
        \addr Centre for Addiction and Mental Health, Toronto, CAN\\
       \addr Department of Statistics, North Carolina State University, Raleigh, USA \\
       \addr Department of Statistics, University of Toronto, Toronto, CAN
       \AND
       \name Desmond Zeya Chen \email \\
       \addr Centre for Addiction and Mental Health, Toronto, CAN\\
       \addr Dalla Lana School of Public Health, University of Toronto, Toronto, CAN
       \AND
       \name David Castle \\
       \addr School of Psychological Sciences, University of Tasmania, Hobart, AUS
       \AND Clement Ma \email clement.ma@camh.ca\\
       \addr Centre for Addiction and Mental Health, Toronto, CAN\\
       \addr Dalla Lana School of Public Health, University of Toronto, Toronto, CAN}

\maketitle

\begin{abstract}%
Basket trials can efficiently evaluate a single treatment across multiple diseases with a common shared target. Prior methods for randomized basket trials required baskets to have the same sample and effect sizes. To that end, we developed a general randomized basket trial with an interim analysis (RaBIt) that allows for unequal sample sizes and effect sizes per basket. RaBIt is characterized by pruning at an interim stage and then analyzing a pooling of the remaining baskets. We derived the analytical power and type 1 error for the design. We first show that our results are consistent the prior methods when the sample and effect sizes were the same across baskets. As we adjust the sample allocation between baskets, our threshold for the final test statistic becomes more stringent in order to maintain the same overall type 1 error. Finally, we notice that if we fix a sample size for the baskets proportional to their accrual rate, then at the cost of an almost negligible amount of power, the trial overall is expected to take substantially less time than the non-generalized version.
\end{abstract}


\section{Background}
\subsection{Basket Trials}
With an ever-growing understanding of the biological principles that govern diseases and with the constant need to release useful treatments at a faster rate, basket trials allow scientists to simultaneously evaluate a single treatment across the biological processes underpinning multiple disorders to efficiently study each disorder/treatment together, rather than individually \citep{woodcock_master_2017, renfro_statistical_2017}. Given the development of biomarkers and genomic identification, basket trials have found a strong foothold in oncology clinical trials, typically in phase 2 trials as exploratory or single-arm trials \citep{hobbs_basket_2022, Park_Siden_Zoratti_Dron_Harari_Singer_Lester_Thorlund_Mills_2019}. In these trials, different tumour subtypes with a common target are typically separated into different baskets, and all participants are given a common therapeutic agent to test for the overall efficacy of that agent. 

There is also interest in conducting randomized basket trials for late phase development \citep{kasim_basket_2023}. In these trials, participants are randomized within each basket to the experimental or control arms, and the results from each tumour subgroup/basket are pooled together (a pooled analysis), enabling simultaneous inference on all diseases and their biomarkers, even those which might be rare in the population. Molecularly targeted therapy based on tumour molecular profiling versus conventional therapy for advanced cancer (SHIVA) \citep{tourneau_molecularly_2015} is one example of a modified basket trial used to test a precision medicine plan. Individuals were enrolled based on a mutation in the hormone receptor, PI3K/AKT/mTOR, or RAF/MEK biological pathways, constituting three baskets. All individuals were randomized to the control (physician’s choice) or treatment (pathway targeting treatment plan) group, but the trial was analyzed using a pooled analysis of progression-free survival.  
In mental health, a randomized basket trial was employed in the phase-3 investigation of Pimavanserin for dementia-related psychosis \citep{Tariot_2021}. With the common treatment, individuals were placed in different baskets based on their type of dementia (Alzheimer's dementia, Parkinsons Disease, Dementia with Lewy Body, frontotemporal, or vascular). This is in contrast to the SHIVA trial which was able to directly use biological pathways, but because mental disorders don’t have well understood genetic biomarkers, we can use common symptoms or disease etiology. Similarly to the SHIVA design, the results from each basket were pooled together for analysis. However, both of these trials illustrate simple examples of basket trials, but numerous improvements to these designs have been made in order to improve the power of the original basket trial.

\subsection{ Overview of Some Basket Trial Methods}

Statistical advancements in both exploratory and randomized basket trials have been made. Some examples of developments in exploratory basket trials have been heavily underpinned with a Bayesian framework. Basket trials using Bayesian hierarchical models have been developed \citep{berry_bayesian_2013} to add an underlying prior distribution on the response rates of all the baskets. Further advancements have altered the information sharing involved in these designs to ensure that not too much information is shared in case the modality of the prior doesn't match the modality of the actual response rates \citep{chu_bayesian_2018}. The idea of information sharing extended out of hierarchical models into model-averaging methods \citep{psioda_bayesian_2021} where each basket is allowed to share information with baskets similar to itself, preventing active baskets from sharing information with inactive baskets and vice-versa. Extensions into machine learning have also been used in hopes of removing the assumption of basket homogeneity by using more complicated models to find subgroups within patients rather than within baskets \citep{pan_bayesian_2022}. Conversely, little has been done with respect to phase III basket trials \citep{kasim_basket_2023} beyond the existing pooled analysis used in SHIVA and Pimavanserin trials along with the designs proposed by \cite{chen_statistical_2016}. Non-phase III randomized designs have also been proposed, one example being a phase 2 design for progression-free survival with multiple arms \citep{Sun_Chen_Patel_2009}.  

\cite{chen_statistical_2016} proposed a phase 3 basket trial to improve power over standard basket trials. Chen et al. kept techniques such as pooling to assess all the baskets together but added an interim analysis where inactive baskets (baskets/diseases/disorders that did not respond to the treatment) were pruned away for futility. The pruning step proposed by Chen et al. spent type 1 error at an interim time point to remove inactive baskets from the analyses to make claims about the remaining active baskets. By “spending” type 1 error, or alpha ($\alpha$), at an interim time point, a new alpha threshold for the final pooled analysis was derived to maintain an overall type 1 error rate. Derivations for the new type 1 error rate for the final pooled analysis ($\alpha^*$) were calculated in Chen et al. based on the desired type 1one error rate at the interim analysis ($\alpha_T$) and the overall study type 1 error rate ($\alpha$). Moreover, Chen et al. proposed three resampling strategies for their design. The first had no resample procedure; the second increased the sample size of the remaining baskets after pruning so that the initial planned sample size was the sample size of the pooled analysis, and the third resampled individuals so that the total sample size was fixed at the initial planned sample size (if a basket was pruned, then only the remaining sample size of that basket would be reallocated). Chen et al. concluded that the second, called “D2”, was the most powerful resampling choice and more powerful than conducting separate trials for each basket. The main condition for this design, however, was that all baskets were equally sized, and calculations for power assumed that the expected standardized effect size for each basket was the same. It is this “D2” method that we expand upon in the strategy reported here

\subsection{Motivating Trial}
Obsessive-compulsive disorder (OCD) is generally considered in a class of mental health disorders called obsessive-compulsive-related disorders (OCRDs), characterized by repetitive, unwanted thoughts or fears that are symptomatically expressed as repetitive behaviors \citep{Fornaro_Gabrielli_Albano_Fornaro_Rizzato_Mattei_Solano_Vinciguerra_Fornaro_2009}. Symptomatically similar is body dysmorphic disorder (BDD), a type of OCRD. People diagnosed with BDD have thoughts related to misconceived defects in their physical appearance that are acted out, inter alia, through repetitive behaviours (eg. mirror checking) or avoidances (eg. of social situations, or ones where they believe their ‘defect’ will be noticed and judged by others). Research in connections and divergences \citep{twenge_age_2019} between OCD and BDD further links them and discusses the need to inspect them more to gauge their biological similarities. Anorexia nervosa (AN), while not an OCRD, is a type of eating disorder (ED) that has parallels with BDD. Patients diagnosed with AN typically have overvalued beliefs about their body image and weight that symptomatically persist as disturbed eating habits. Similarities between BDD and AN at a neurological and behavioural level have been shown in previous research \citep{phillipou_direct_2019}. 

With increasing efforts to work towards clinical solutions in mental health, it is becoming evermore important to improve the design of confirmatory mental health clinical trials so that effective interventions can be released to the public efficiently. With psychedelics affecting the serotonin systems in the body, it becomes a new area of interest as a possible clinical intervention. Based on the assumption that all three disorders (AN, BDD, and OCD) are linked, it becomes necessary to test how psychedelics, specifically psilocybin, might be used to treat these disorders \citep{ledwos_therapeutic_2023}. 

Most current clinical trials involving psilocybin are either Phase 1 or Phase 2 assessing efficacy and adverse events. Psilocybin is being tested for a range of disorders, including substance abuse \citep{university_of_wisconsin_madison_safety_2023, fink-jensen_psilocybin_2023}, post-traumatic stress disorder \citep{davis_examining_2023}, depression \citep{weintraub_psilocybin-assisted_2023, university_of_chicago_open_2023, usona_institute_randomized_2023}, OCD \citep{johns_hopkins_university_effects_2023}, anxiety \citep{woolley_psilocybin_2024, downar_psilocybin_2024} and bipolar disorder \citep{woolley_open-label_2023}. Given that psilocybin is currently being used to treat a multitude of disorders, there is a need for master protocols to make trials more efficient. Such trial designs could help improve the speed of determining a clinical effect and establishing which indications to pursue, and which to drop. 

\subsection{Study Objective}
To address these issues, we propose a basket trial to simultaneously investigate the effects of psilocybin across obsessive-compulsive disorder, body dysmorphic disorder, and anorexia nervosa using the most powerful “D2” method developed by Chen et al. However, Chen’s design requires equal-sized baskets with the same effect sizes. This proves to be a limitation in two ways. In the case that one basket might be slower to accrue, varying the planned basket sizes could improve the speed of the trial. Moreover, being able to account for different anticipated effect sizes could lower the required number of participants to have a well powered study.

Thus, we propose the Randomized Basket design with an Interim Analysis (RaBIt) to generalize the method to allow baskets of different sizes and different anticipated effect sizes per basket. We calculate the power of such a design and simulate the duration of the trial under different conditions. 
\section{Methods}
RaBIt is described in four steps. In the first step, we outline the notation and definitions of the variables, parameters, and statistics. Next, we describe the key differences in the formula between RaBIt and "D2" by Chen et al. Thirdly, we present all the formulae for calculating power and sample size. Lastly, we describe the methods used to investigate RaBIt.

\subsection{Extensions and Definitions from Chen et al.}
We expand on the notation from  Chen et al. Let $K$ be the number of baskets, and $N$ be the total number of participants in the study. We denote $\mathbf{p}$ as the proportion vector, where $p_i$ represents the proportion of the total population, $N$, that is initially allocated to basket $i$ before any baskets are pruned. Consequently, $p_i \cdot N$ represents the total number of individuals initially allocated to the $i$th basket. Let $\mathbf{m}$ be the indicator vector of length $K$ where $m_i =0$ represents that the $i$th basket was pruned away at the interim analysis, while $m_j=1$ represents that the $j$th basket was not pruned at the interim analysis. Next, let $\mathcal{M} = \{0,1\}^K$ denote the set of possible combinations of baskets that make it past the interim; this means that $\mathbf{m}\in\mathcal{M}$ for all possible combinations of 0's and 1's of length $K$. We then let $t$ represent the information time; consequently, $N \cdot p_i \cdot t_i$ represents the total number of participants recruited for the interim analysis for the $i$ th basket tested at a threshold of $\alpha_t$ (interim analysis type 1 error rate). Let $\mathbf{\Delta}$ represent the vector of effect sizes where $\Delta_i$ represents the (anticipated) effect size of the $i$th basket. 

Next, let $Y_{i1}$ represent the standardized interim analysis test statistic for the trial where under the null hypothesis, $Y_{i1} \sim \mathcal{N}(0,1)$, or under the alternative hypothesis $Y_{i1} \sim \mathcal{N}(\Delta_i\sqrt{(N \cdot p_i \cdot t_i)/4}, 1)$ when considering a 1:1 randomization within baskets. We then let $\mathbf{g}$ be the indicator vector (constructed similarly to $\mathbf{m}$) to indicate which baskets are truly active under the alternative hypothesis. Next, we described $\mathbf{j}$ to be the element-wise product of $\mathbf{m}$ and $\mathbf{g}$ which represents which baskets are both active and make it past the interim analysis. We also introduce a function $id(\cdot):\{0,1\}^K \rightarrow \{1,2,...,K\}$ which takes in an indicator vector (either $\mathbf{m,g,j}$) and returns a set about which baskets are valued at 1. Lastly, we let $|\cdot |$ denote the cardinality of a set. For example, $|id(\bm{m})|$ denotes the number of baskets that make it past the interim analysis.

In Chen et al.'s "D2" method, the total sample pruned away from the number of baskets removed is reallocated equally to the remaining baskets. Alternatively, we defined weights, $w_i$, to represent the proportion of the total sample size that the $i$th basket after inactive baskets are pruned at the interim analysis; with the goal of maintaining efficiency from having different sized baskets, we proportionally allocated the pruned sample to the remaining baskets based on the respective size of the remaining baskets.

\begin{equation}
    w_i = \frac{p_i}{\mathbf{m}\cdot \mathbf{p}}: i\in id(\mathbf{m}) \label{weight}
\end{equation}

The denominator is the dot product of the proportionality and interim-active vectors representing the total proportion of the sample left after pruning, and the numerator is the proportion originally allocated to the $i$th basket. This ensures that after the interim analysis, the remaining baskets are proportionally equivalent with respect to how they were originally set. Consequently, this results in the fact that $Y_{i2} \sim \mathcal{N}(0,1)$ under the null hypothesis, or under the alternative hypothesis $Y_{i2} \sim \mathcal{N}(\Delta_i\sqrt{(N \cdot w_i)/4}, 1)$.

Next, we generalize how the final pooled test statistic was calculated. Considering each basket now has a different size, we construct a test statistic $V_\mathbf{m}$  (tested at a significance level of $\alpha^*$) using the Weighted Stouffer's Z-Score Method \citep{Stouffer_Suchman_Devinney_Star_Williams} (as opposed to the unweighted one used in Chen et al.).

\begin{equation}
    V_{\bm{m}} = \frac{\sum_{i \in id(\bm{m})} w_i Y_{i2}}{\sqrt{\sum_{i \in id(\bm{m})} w_i^2}} \sim \mathcal{N}(0,1)
    \label{teststat}
\end{equation}

\begin{figure}[H]
\centering
\includegraphics[width=\textwidth]{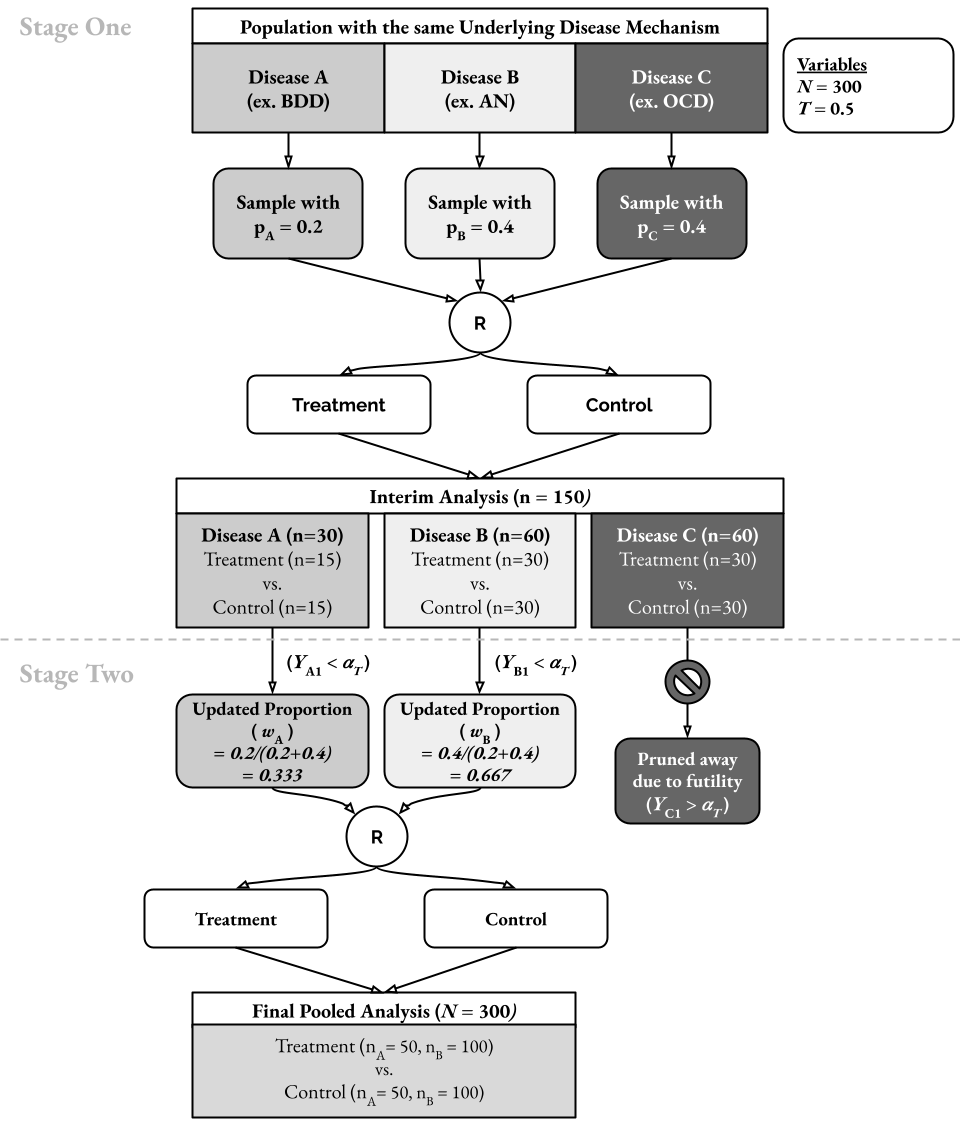}
\caption{An example generalized basket trial design allowing for the baskets to be a different size. In this example, the third basket is pruned away, and the remaining baskets are scaled accordingly.}
\label{exampledesign}
\end{figure}

Overall, a 3 basket example of this trial can be visualized in Figure \ref{exampledesign}.

\subsection{Maintaining an Overall Type 1 Error Rate}
As in Chen et al., to determine the power of our final pooled test statistic, we first need to maintain the overall type 1 error at $\alpha$. Given the prespecified interim type 1 error rate of $\alpha_t$, an $\alpha^*$ threshold is needed for the final pooled analysis so that an overall $\alpha$ type 1 error rate is maintained. To calculate $\alpha^*$, we calculate the probability of the final pooled test statistic being significant under the null hypothesis given a choice of $\bm{m}, \alpha_t$, and $\alpha^*$, from this, we can use a root solver to solve numerically the value of $\alpha^*$. We began with the probability of $V_\mathbf{m}$ being significant under the null hypothesis (none of the groups are active). This is the extension of equation 2 and 2a from Chen et al.

\begin{align}
    &\Pr_{H_0}(V_{\bm{m}}|\alpha^*, \alpha_t, \bm{m}, \bm{\Delta})\\
    &= \Pr_{H_0}\left(\bigcap_{i \in id(\bm{m})}Y_{i1} > Z_{1-\alpha_t},\bigcap_{l \not\in id(\bm{m})}Y_{l1} > Z_{1-\alpha_t}, V_{\bm{m}} > Z_{1-\alpha^*} \right)\\
    &= \Pr_{H_0}\left(\bigcap_{i \in id(\bm{m})}Y_{i1} > Z_{1-\alpha_t}, V_{\bm{m}} > Z_{1-\alpha^*} \right) \cdot (1-\alpha_t)^{K-|id(\bm{m})|}
    \label{null_error}
\end{align}

We noted that the above joint probability in the solution to \ref{null_error} requires the covariance between our interim test statistic $Y_{i1}$ and $V_{\bm{m}}$. We derive the covariance as:

\begin{align}
    corr(Y_{i1}, V_{\bm{m}}) &= corr(Y_{i1}, Y_{i2}) \cdot corr(Y_{i2}, V_{\bm{m}})\text{ where}\\ 
    corr(Y_{i1}, Y_{i2}) &= \sqrt{t(\bm{m}\cdot\bm{p})}\label{first_corr}\\
    corr(Y_{i2}, V_{\bm{m}}) &= \frac{w_i}{\sqrt{\sum_{i=1}^m w_i^2}}\label{sec_corr}\\
    \implies corr(Y_{i1}, V_{\bm{m}}) &= \frac{w_i}{\sqrt{\sum_{i=1}^m w_i^2}}\sqrt{t(\bm{m} \cdot \bm{p})}
    \label{final_corr}
\end{align}

Where the result of equation \ref{first_corr} comes from the fact that the initial test statistic sample size is $t\cdot p_iN$, but it increases to $w_iN = \frac{p_i}{\bm{m}\cdot \bm{p}}$ implying an increase of the basket's sample size by a factor of $\frac{1}{(\bm{m}\cdot \bm{p})t}$. Equation \ref{sec_corr} is determined by the weights of in equation \ref{teststat}. 

The overall type 1 error $\alpha$ is given by summing over every choice of $\bm{m}$. 

\begin{equation}
    \alpha = \sum_{\bm{m} \in \mathcal{M}}\Pr_{H_0}(V_{\bm{m}}|\alpha^*, \alpha_t, \bm{m}, \bm{\Delta})
    \label{alphaeq}
\end{equation}

Then, we can use a root solver on equation \ref{alphaeq} to calculate the appropriate value of $\alpha^*$ to maintain the overall type 1 error of the trial (equivalent to equation 3 in Chen et al.)

The major difference between the generalization and equation 3 from Chen et al. is the change from combinatorics to just a simple sum. This is a consequence of allowing each basket to have different parameters (like effect size or proportion), preventing the simplification in the calculation done by Chen et al., which takes advantage of the fact that the probabilities are the same when the same number of baskets make it past the interim analysis.

\subsection{Calculating Power}
For calculating power, we establish the alternative hypothesis $H_{1\bm{g}}$ that some choice of baskets, $\bm{g}$, are truly active, but we calculate the power for the final pooled test statistic $V_{\bm{m}}$ being significant. Note that this means that we calculate the power for the final pooled analysis rather than for just the specific $\bm{g}$ baskets being the only ones included in $V_{\bm{m}}$. For example, truly non-active baskets could be included or truly active baskets missing from the final pooled statistic  calculation. For simplicity of calculations we generalize the distributions of both of the statistics to the ones found below assuming a 1:1 randomization within baskets:

\begin{align*}
    Y_{i1}&\sim \mathcal{N}(g_i\Delta_i \sqrt{p_iNt\cdot/4}, 1)\\
    Y_{i2}&\sim \mathcal{N}(j_i\Delta_i\sqrt{(N \cdot w_i)/4}, 1)\\
    \implies V_{\bm{m}}&\sim \mathcal{N}\left(\sum_{j \in id(\bm{j})}\frac{w_j}{\sum_{m \in id(m)}w_m^2}\Delta_j\sqrt{\frac{N}{4}}\sqrt{w_j},1\right)
\end{align*}

Thus in order to calculate the power of $V_{\bm{m}}$, we start by calculating the probability it is significant under some fixed choice of $\bm{m}$ (similar to equation 4 in Chen et al.).

\begin{align}
&\Pr_{H_{1\bm{g}}}(V_{\bm{m}}|\alpha^*, \alpha_t, \bm{m}, \bm{\Delta})\notag \\
=& \Pr_{H_{1\bm{g}}} \left(\bigcap_{i\in id(\bm{m})}Y_{i1} \geq Z_{1-\alpha_t}, \bigcap_{j\notin id(\bm{m})}Y_{j1}<Z_{1-\alpha_t}, V_{\bm{m}} \geq Z_{1-\alpha^*} \right)\notag\\
=& \Pr_{H_{1\bm{g}}} \left(\bigcap_{i\in id(\bm{m})}Y_{i1}\geq Z_{1-\alpha_t}, V_{\bm{m}} \geq Z_{1-\alpha^*} \right)
 \Pr_{H_{1\bm{g}}} \left(\bigcap_{j\notin id(\bm{m})}Y_{j1}<Z_{1-\alpha_t} \right)\quad
\label{singular_power}
\end{align}

Where the first probability statement in equation \ref{singular_power} models what was found in equation \ref{null_error} using the correlation result found in equation \ref{final_corr}, and the second statement can be rewritten for simplicity as it represents the baskets that were pruned at the interim. To do this, we note that this would be the product of baskets accurately getting pruned away (let there be $R$ of them) and baskets inaccurately getting pruned away.

\begin{align*}
    &\Pr_{H_{1\bm{g}}}\left(\bigcap_{i \not\in id(\bm{m})} Y_{i1}<Z_{1-\alpha_t}\right)\\ 
    &= \left(\prod_{i \in id(\bm{g})\setminus id(\bm{j})} \Phi(Z_{1-\alpha_t} -\Delta_i\sqrt{p_iN_i/4})\right) \cdot (1-\alpha_t)^{R}
\end{align*}
    
Thus the power of the final pooled test statistic, $V_{\bm{m}}$, can be solved by summing over every possible scenario of $\bm{m} \in \mathcal{M}$ baskets making it past interim. We note that if no baskets make it past interim, the power of the final pooled test statistic is inherently 0. Thus we arrive at the final formulation for power of the final pooled test statistic for this trial that mimics equation 5 in Chen et al.

\begin{equation}
    1 - \beta =  \sum_{\bm{m} \in \bm{\mathcal{M}}}\Pr_{H_{1\bm{g}}}(V_{\bm{m}}|\alpha^*, \alpha_t, \bm{m}, \bm{\Delta})
\end{equation}

\subsection{Gini Impurity}
To evaluate RaBIt, we use the Gini Impurity to measure the level of equality between baskets.

\begin{equation}
    \text{Gini Impurity} = 1 - \sum_{i=1}^Kp_i^2
\end{equation}

The higher the Gini Impurity, the more equal the sample size is between baskets, while a smaller Gini Impurity would indicate that a few baskets hold a majority of the sample size (have a higher proportion or weight) than the remaining baskets.

\subsection{Estimators for Trial Duration and Realized Sample Size}
To estimate the expected trial duration and realized sample size, let $\bm{A}$ be the accrual vector such that $A_i$ is the number of people accrued to basket $i$ per unit of time. The expected trial duration under a choice of $\bm{A}$, $E(D_{\bm{A}})$, is calculated by noting the trial duration is specified by the number of baskets that make it past the interim; thus the expected value of the trial duration is the product of the trial duration (when $\bm{m}$ baskets make it through the interim with some choice of $\bm{A}$), $d_{\bm{A},\bm{m}}$, and the probability that trial duration occurs (the probability of that specific $\bm{m}$ vector being realized in the trial), $\Pr_{H_{1\bm{g}}}(\bm{m})$, summed over every possible way the trial could be conducted (all $\bm{m} \in \mathcal{M}$).

\begin{equation}
    E(D_{\bm{A}}) = \sum_{\bm{m} \in \mathcal{M}} \Pr_{H_{1\bm{g}}}(\bm{m}) \cdot d_{\bm{A},\bm{m}}
\end{equation}

Given that each group standardized test statistic at the interim analysis is assumed to be independent, $\Pr_{H_{1\bm{g}}}(\bm{m})$ reduces down to the product of the specified baskets making it past the interim analysis and the others being pruned away under the alternative hypothesis.

\begin{equation}
    \Pr_{H_{1\bm{g}}}(\bm{m}) = \Pr_{H_{1\bm{g}}}\left(\bigcap_{i\in id(\bm{m})}Y_{i1}\geq Z_{1-\alpha_t}, \bigcap_{i \not\in id(\bm{m})} Y_{i1}<Z_{1-\alpha_t}\right)
    \label{expectation_prob}
\end{equation}

Deriving a specific value of trial duration $d_{\bm{A},\bm{m}}$ is more complicated. Given the overall goal of adding different-sized baskets in order to match the accrual rate of each basket (especially assuming they are constant over time), we estimate $d_{\bm{A},\bm{m}}$ using the fastest possible, though impractical, design provided in the supplementals.

The expected number of participants $E(P)$ can also be calculated after noting that the number of participants in the trial is determined by which trials make it past the interim. This is a consequence of the final pooled statistic having a sample size of $N$, then any extension beyond $N$ is a result of baskets being pruned. Thus we get the following expectations for the number of participants:

\begin{align*}
    E(P) &= \sum_{\bm{m} \in \mathcal{M}} \Pr_{H_{1\bm{g}}}(\bm{m}) \cdot P_{\bm{m}}\\
    \text{where }  P_{\bm{m}} &= N + \sum_{i \notin id(\bm{m})}Np_it
\end{align*}

Where the above uses the result from equation \ref{expectation_prob}. Moreover, both expectations are built off of discrete random variables (based on the number of baskets that make it past the interim), and as a result, the variance and corresponding confidence interval of these estimates can be made.

\subsection{Validation and Evaluation of RaBIt}
We generalized the code originally published by Chen et al. To validate the code, a series of tests were done using all combinations of $K=2,3,6$ baskets with $N=300, 420$ total participants, standardized effect sizes of $\Delta_i = 0.2, 0.5$, and information times $t=0.3, 0.5$ with each basket receiving an equal initial proportion of the sample size. Power is calculated using the original code provided by Chen et al., and the code that has updates made to account for the generalization. This is done to ensure that our results are consistent with the "D2" methodology provided by Chen et al.

Then across $K=2, 3$ baskets, with an interim time $t_i=0.5$, and an interim alpha $\alpha_t=0.3$, controlling with $\alpha =0.025$, with a planned total sample size of $N=150$, and effect size of $\Delta_i=0.5$ for each basket, $\alpha^*$ and power is solved for every possible version of participant allocation. These results are then plotted against their Gini Impurity. It is important to note that by attributing a $\Delta_i \neq 0$ value to each basket, we are concurrently saying it is truly active under the alternative hypothesis. 

To understand how having different anticipated effect sizes between baskets alters the overall power of the trial, a 3-basket design is used using the same parameters described above. However, 4 scenarios for which the average effect size is 0.2 and 0.5 are conducted where the effect size diverges from equality across each basket. Power is then tabulated for each scenario.

Lastly, the expected trial duration, power, expected number of participants, worst case number of participants, and 95\% confidence intervals for both expected duration and participants are calculated for a 3-basket trial using the parameters discussed previously (fixing the effect size of each basket $\Delta = 0.5$). Accrual vectors are tested with $\bm{A}=(2,2,2), (2,2,1), (3, 1,1)$, and proportion vectors starting at basket equality (as in the case of Chen et al.), moving towards basket size proportional to the accrual rate, and one more example beyond.

\subsection{Software}
Software is provided in the supplemental materials. An interactive, web-based R Shiny implementation of RaBIt is available online \citep{Chen_Patel_Xie_Chen_Castle_Ma}.
\section{Results}
\subsection{Validation}
To ensure consistency of RaBIt with "D2" power was calculated across a variety of designs (altering effect and sample sizes, along with the number of baskets) with equal allocation of the sample to each basket. As shown in Table \ref{ConsitencyResults}, power values between "D2" and RaBIt only deviates $\pm 0.3 \%.$

\subsection{Effect of Basket Allocation on $\alpha^*$}
\begin{figure}[H]
\centering
\includegraphics[width=\textwidth]{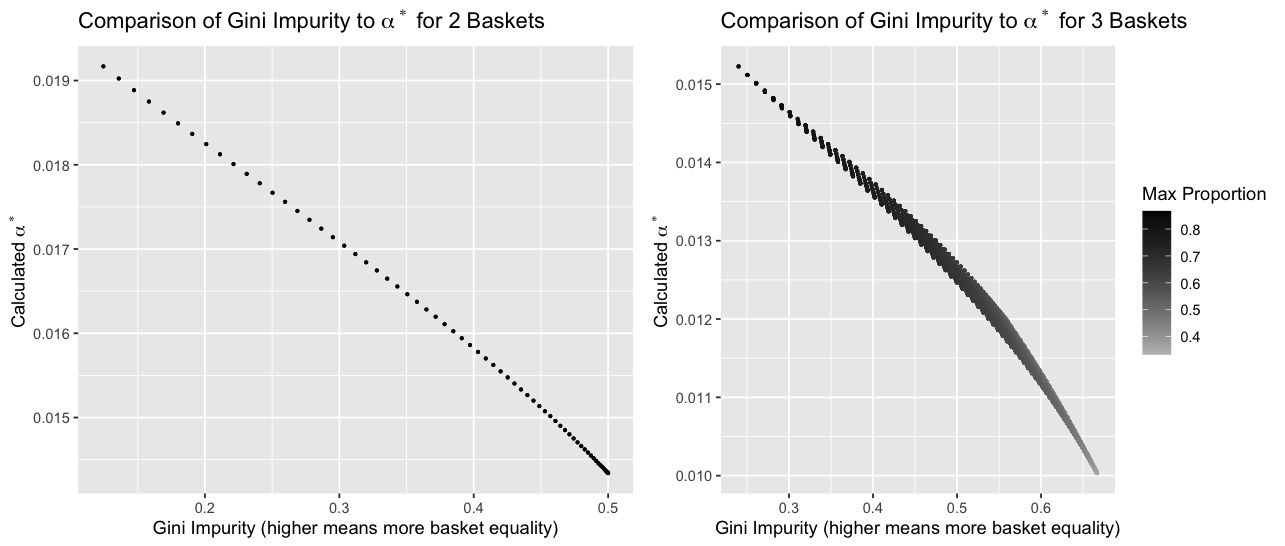}
\caption{Using a sample size of 150, effect size of 0.5, interim time of 0.5, controlled type 1 error rate of 0.025, interim type 1 error rate of 0.3, the solved $\alpha^*$ is plotted against the gini impurity for both a two (left) and three basket trial (right).}
\label{alphastarfig}
\end{figure}

To evaluate the effect of sample size allocation on the interim significance threshold,  $\alpha^*$, we calculated $\alpha^*$ for every possible participant allocation for $K=2,3$ baskets (with a minimum of 10 participants in a basket for practicality). In Figure \ref{alphastarfig}, as the Gini Impurity increases (e.g. basket sizes are more equal), the $\alpha^*$ threshold becomes stricter. For K=2 baskets, the $\alpha^*$ is lowest with equally sized baskets (75 participants per basket) at $0.0143$, and $0.0192$ with the most unequally-sized baskets  (10 and 140 per basket). In K=3 baskets, $\alpha^*=0.0100$ when the basket sizes are equal (50 per basket), and $\alpha^*=0.0152$ for the most unequal basket allocation tested (10, 10, and 140 per basket). 

\subsection{Proportions versus Power}
\begin{figure}[H]
\centering
\includegraphics[width=\textwidth]{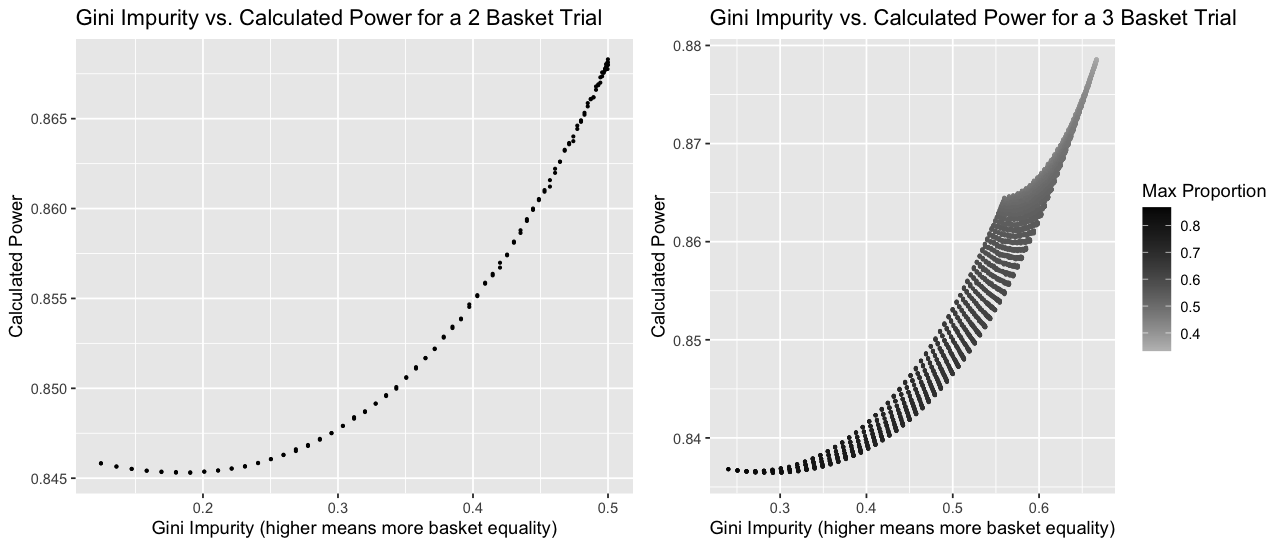}
\caption{Using a sample size of 150, effect size of 0.5, interim time of 0.5, controlled type 1 error rate of 0.025, interim type 1 error rate of 0.3, power plotted against the gini impurity for both a two (left) and three basket trial (right).}
\label{powerfig}
\end{figure}

 In comparison, increasing the Gini Impurity slightly decreases the overall power of the final pooled analysis (Figure \ref{powerfig}). For K=2 baskets, power differed by 0.022 (0.868 vs. 0.846) for equal allocation vs. the most unequal allocation, respectively. For K=3 baskets, power differed by 0.042 (0.879 vs. 0.837).

\subsection{Effect Size Versus Power}
Next, we examined the effect of altering the effect size per basket for a fixed sample size of 150 participants (equal allocation). As the average effect size increases, so does the power of the final test statistic (Table \ref{ESResults}). Moreover, as the effect size of one basket increases and the average effect size is maintained, the power for the final pooled test statistic also increases. For example, with an average standardized effect size of 0.5, as the largest effect size basket goes from 0.5 to 1.1, the power increases from 0.879 to 0.969 for the final pooled test statistic.

\subsection{Effect of Basket Allocation on Trial Duration}
Finally, we examined the trial duration and the expected number of participants for 3 scenarios with different accrual rates and proportion vectors. In Table \ref{TimeVec}, as the proportion vector becomes closer to being proportional to the accrual rate, the expected duration of the trial is much lower than all other cases. For example, for a 3 basket trial with accrual rates of $A=(2,2,1)$ participants per month, the trial duration is shortest (43.58 months) when the allocation proportion is proportional to the accrual rates (e.g. 40\%, 40\%, and 20\% for baskets 1, 2, and 3, respectively). In comparison, equal allocation requires a longer trial duration (61.12 months).  

Moreover, the difference in power from these two designs is approximately 0.0025. This means that at an almost negligible loss in power, allocating participants proportional to the expected accrual rates can reduce the duration of a trial by 17.54 months in this uniformly accruing example. Moreover, when looking at the upper bound (if faster-accruing baskets are pruned away) of the 95\% confidence interval for the trial duration, we notice the generalization reduces the 95\% upper confidence limit from 77.36 to 59.25 months.

\section{Discussion}
\subsection{Interpretation of Results}
RaBIt improves the utility of Chen’s two-stage randomized basket design by allowing baskets with unequal sample sizes and effect sizes. Overall, as basket sizes become more imbalanced, power decreases slightly and the final significance threshold becomes less stringent. We believe this is due to a lower chance of making an error in pruning baskets during the interim analysis. As a basket receives a greater proportion of the total sample size, its interim sample size and thus, interim power is increased, meaning a smaller likelihood of making an error at the interim. Conversely, baskets with smaller sample sizes have low power at the interim analysis and are likely to be pruned away. Thus, it is more likely that only the larger baskets remain for the final analysis, and requires a less stringent threshold for the final pooled analysis. 

Such a phenomenon also explains why a singular basket's anticipated effect size affects the design's overall power. A singular basket with more power in the interim gives it a lower chance of being falsely rejected, thus giving a significant final pooled test statistic. Lastly, the power of the pooled final test statistic decreases for designs with greater imbalance. However, in cases of lower levels of an anticipated effect size, the trial increases in power as one basket starts to dominate the trial. Consequently, this RaBIt design is most applicable when we believe effect sizes are relatively similar. A basket trial designed in this fashion would protect against pruning the low-effect baskets away. Practically, if specific baskets (indications) are known to have small effect sizes, investigators should consider completely excluding these baskets from the trial.

When the basket sizes are proportional to the accrual rate, the trial is expected to have the shortest duration. Trial duration increases when basket sizes are not proportional to the accrual rate since the trial will need to remain open to enroll to baskets with slower accrual rates. Thus, RaBIt can shorten trial duration compared to Chen’s design by allowing basket sizes to match the (unequal) prevalence of the different disease/disorders/biological pathways and their accrual rates. Moreover, as the proportion vector is altered away from basket equality, there is an almost negligible loss in power for the final pooled statistic in comparison to the case where baskets are sampled equally. However, in a phase 3 trial, it is important to not undersample any basket. A very small-sized basket would be almost guaranteed that be pruned away. Investigators can decide upon minimum basket sizes by calculating the power desired for the within-basket interim analyses. 

\subsection{Application in the Motivating Trial}
We applied RaBIt to our motivating trial of psilocybin for obsessive-compulsive disorder, body dysmorphic disorder, and anorexia nervosa. Given the clinical connections between the disorders and the belief in a common biological mechanism, a basket trial is a reasonable design to simultaneously test psilocybin against the common pathway. From prior clinical experience, we expect BDD to have half the accrual rate as OCD and AN. Thus, we propose a generalized basket trial with a proportion vector $\bm{p} = (0.2, 0.4, 0.4)$ for the BDD, OCD, and AN baskets respectively. As shown in Table \ref{TimeVec}, a sample size of 150 participants achieves 87\% power for the study, assuming a standardized effect size 0.5 per basket. However, sample sizes can be recalculated after concluding the Phase 2 study and getting a better estimate of effect sizes per basket.

\subsection{Applications Beyond the Motivating Trial}
While we focused primarily on the application of RaBIt in mental health, this design is easily applicable to other types of clinical trials. Using basket trial designs in oncology using biomarkers are not uncommon \citep{kasim_basket_2023}. Beyond mental health and oncology, basket trials are beginning to be used when testing therapies against neurodegenerative disorders \cite{cummings_role_2022} as well in schizophrenia trials using biomarkers to define baskets \citep{Joshi_Light_2018}. Overall, this design applies to any clinical investigation between a single intervention and multiple diseases/disorders/subgroups (baskets) where there is a believed biological connection between the baskets. This simple phase 3 basket trial design offers a frequentist framework to investigate multiple diseases simultaneously, building upon already existing methods that use pruning and pooling steps in their analysis. While there is no information sharing, as common in the Bayesian designs, this basket trial method offers a type 1 error-controlled way to test an intervention without having to set any priors.

\subsection{Limitations}
Our study has several limitations. First, we only examined the  "D2" sample size re-allocation strategy proposed by Chen et al. and did not generalize other methods. We focused on the “D2” method since Chen demonstrated that it had the highest power. Second, we did not examine the power or sample size for designs with different endpoints at the interim and final analysis that might be used for early stopping for oncology trials. Finally, although we performed a comprehensive evaluation of designs with different basket sizes ($\bm{p}$), effect sizes ($\bm{\Delta}$), number of baskets ($K$), and accrual rates ($\bm{A}$), we did not vary the interim significance threshold ($\alpha_T$), overall type 1 error rate ($\alpha$), information time ($t$), and randomization ratio.

\subsection{Conclusions}
Our generalized two-stage randomized basket trial with an interim analysis (RaBIt) provides a way to match basket size with the expected accrual rate of each basket to provide a more efficient trial with a nearly negligible loss in power. Originally motivated by the anticipated phase III investigation of psilocybin against OCD, BDD, and AN, this generalization is a generalized version of the D2 method proposed by Chen et al. that prunes inactive baskets at an interim point and pools the baskets at the end for the final analysis and resamples into the at-interim-active baskets in a novel way that has been shown to improve the efficiency of the "D2" trial originally laid out by Chen et al. However, while we argued that such a design is applicable in mental health clinical trials, RaBIt is applicable in any investigation between multiple diseases/disorders/subgroups where there is believed to be an underlying biological connection to the disorders being studied against a single intervention. While this generalization is a step towards building more efficient clinical trials, future directions could include multiple arms per basket, a shared control group, and more. However, the current trial design provides an efficient way to evaluate multiple disorders against a single treatment in a way that incorporates detailed external information (proportions to match accrual rates or different anticipated effect sizes) to improve the efficiency of the trial.

\acks{All acknowledgements go at the end of the paper before appendices and references.
Moreover, you are required to declare funding (financial activities supporting the
submitted work) and competing interests (related financial activities outside the submitted work).
More information about this disclosure can be found on the JMLR website.}


\newpage

\appendix
\section{Tables and Figures}
\begin{table}[H]
\small\sf\centering
\caption{Power of the equal proportion design calculated with both the original code from Chen et al. and from the updates made with the generalization across various designs in order to ensure consistency in the results}
\begin{tabular}{c c c c c c }
\toprule
Number & Samples & 
Standardized & Information & Power & Power\\
of Baskets & per Basket & Effect Size & Time & (Chen) & (Generalization)\\
\midrule
 2 & 150 & 0.2 & 0.3 & 0.4079576 & 0.4079349 \\
 2 & 150 & 0.2 & 0.5 & 0.4017306 & 0.4017735 \\
 2 & 150 & 0.5 & 0.3 & 0.9883213 & 0.9883675 \\
 2 & 150 & 0.5 & 0.5 & 0.9878961 & 0.9876536 \\
 2 & 210 & 0.2 & 0.3 & 0.5317049 & 0.5316057 \\
 2 & 210 & 0.2 & 0.5 & 0.5231576 & 0.5232334 \\
 2 & 210 & 0.5 & 0.3 & 0.9985419 & 0.9985403 \\
 2 & 210 & 0.5 & 0.5 & 0.9986655 & 0.9986833 \\
 3 & 100 & 0.2 & 0.3 & 0.4101353 & 0.4101435 \\
 3 & 100 & 0.2 & 0.5 & 0.4027752 & 0.4028885 \\
 3 & 100 & 0.5 & 0.3 & 0.9894267 & 0.9894057 \\
 3 & 100 & 0.5 & 0.5 & 0.9872007 & 0.9870835 \\
 3 & 140 & 0.2 & 0.3 & 0.5345540 & 0.5345404 \\
 3 & 140 & 0.2 & 0.5 & 0.5242972 & 0.5242650 \\
 3 & 140 & 0.5 & 0.3 & 0.9988763 & 0.9988663 \\
 3 & 140 & 0.5 & 0.5 & 0.9986178 & 0.9985386 \\
 6 & 50 & 0.2 & 0.3 & 0.4133663 & 0.4134196 \\
 6 & 50 & 0.2 & 0.5 & 0.4096576 & 0.4096276 \\
 6 & 50 & 0.5 & 0.3 & 0.9905375 & 0.9905409 \\
 6 & 50 & 0.5 & 0.5 & 0.9886482 & 0.9886300 \\
 6 & 70 & 0.2 & 0.3 & 0.5388185 & 0.5388941 \\
 6 & 70 & 0.2 & 0.5 & 0.5331173 & 0.5331435 \\
 6 & 70 & 0.5 & 0.3 & 0.9991008 & 0.9990958 \\
 6 & 70 & 0.5 & 0.5 & 0.9988114 & 0.9987675 \\
\bottomrule
\end{tabular}
\label{ConsitencyResults}
\end{table}
\begin{table}[H]
\small\sf\centering
\caption{Using a sample size of 150, interim time of 0.5, controlled type 1 error rate of 0.025, interim type 1 error rate of 0.3, power is calculated assuming all baskets are active with the given effect sizes per basket.}
\begin{tabular}{c c c }
\toprule
Average Effect Size & Effect Size per Basket & 
Power \\
\midrule
 0.2 & (\textbf{0.2}, 0.2, 0.2) & 0.2454 \\
 0.2 & (\textbf{0.3}, 0.2, 0.1) & 0.2837 \\
 0.2 & (\textbf{0.4}, 0.1, 0.1) & 0.3636 \\
 0.2 & (\textbf{0.5}, 0.05, 0.05) & 0.4868 \\
 \midrule
 0.5 & (\textbf{0.5}, 0.5, 0.5) & 0.8786 \\
 0.5 & (\textbf{0.7}, 0.5, 0.2) & 0.8862 \\
 0.5 & (\textbf{0.8}, 0.6, 0.1) & 0.9428 \\
 0.5 & (\textbf{0.9}, 0.5, 0.1) & 0.9494 \\
 0.5 & (\textbf{1.1}, 0.2, 0.2) & 0.9692 \\
\bottomrule
\end{tabular}
\label{ESResults}
\end{table}

\begin{sidewaystable}
\small\sf\centering
\caption{Using a sample size of 150, interim time of 0.5, controlled type 1 error rate of 0.025, interim type 1 error rate of 0.3, power is calculated assume all baskets are active with the given effect sizes per basket.  The bolded lines have a proportion vector proportional to the accrual rate.}
\begin{tabular}{c c c c c c c c}
\toprule
$\bm{A}$& $\bm{p}$ & $E(D_{\bm{A}})$ & Power & $E(P)$ & 95\% CI for & 95\% CI for Number \\
 & & (months) & & (people) & Time (months) & of Participants (people)\\
\midrule
        (2, 2, 2)&  \textbf{(0.33, 0.33, 0.33)}&  \textbf{36.28}&  \textbf{0.8786}&  \textbf{165.63}&  \textbf{[25.24, 47.32]}& \textbf{[151.77, 179.49]}\\
        &  (0.35, 0.35, 0.3)&  37.63&  0.8785&  165.59&  [26.80, 48.45]& [151.74, 179.44]\\
         &  (0.37, 0.37, 0.26)&  39.23&  0.8778&  165.43&  [28.58, 49.88]& [151.58, 179.27]\\
         &  (0.4, 0.4, 0.2) &  41.58&  0.8760&  164.92&  [31.02, 52.14]& [151.08, 178.76]\\
         &  (0.45, 0.45, 0.1)&  45.29&  0.8717&  163.16&  [34.09, 56.48]& [149.24, 177.07]\\
        \midrule
         (2, 2, 1)&  (0.33, 0.33, 0.33)&  61.12&  0.8786&  165.63&  [44.88, 77.36& [151.77, 179.49]\\
         &  (0.35, 0.35, 0.3)&  56.47&  0.8784&  165.59&  [40.51, 72.44]& [151.74, 179.44]\\
         &  (0.37, 0.37, 0.26)&  51.09&  0.8778&  165.43&  [35.37, 66.80]& [151.58, 179.27]\\
         &  \textbf{(0.4, 0.4, 0.2)}&  \textbf{43.58}&  \textbf{0.8761}&  \textbf{164.92}&  \textbf{[27.90, 59.25]}& \textbf{[151.08, 178.76]}\\
         &  (0.45, 0.45, 0.1)&  46.47&  0.8718&  163.16&  [32.06, 60.89]& [149.24, 177.07]\\
         \midrule
          (3, 1, 1) &  (0.33, 0.33, 0.33)&  68.72&  0.8786&  165.63&  [49.65, 87.79]& [151.77, 179.49]\\
         &  (0.4, 0.3, 0.3)&  62.2&  0.8779&  165.48&  [43.12, 81.28]& [151.64, 179.32]\\
         &  (0.5, 0.3, 0.2)&  57.47&  0.8739&  164.57&  [38.38, 76.56]& [150.83, 178.31]\\
         &  \textbf{(0.6, 0.2, 0.2)}&  \textbf{43.71}&  \textbf{0.8682}&  \textbf{163.44}&  \textbf{[24.36, 63.05]}& \textbf{[149.97, 176.92]}\\
         & (0.7, 0.15, 0.15)& 46.17& 0.8609& 161.57& [29.18, 63.16]&[148.45, 174.69]\\
\bottomrule
\end{tabular}
\label{TimeVec}
\end{sidewaystable}

\newpage

\newpage
\section{Supplementals}

\subsection{Calculating Trial Duration}
To illustrate the advantage of a generalized design, we calculated the shortest possible trial duration for Chen’s designs and the RaBiT design. We assume that the interim analysis will be conducted as soon as a basket reaches it’s target sample size. This strategy may be impractical for trials with many baskets. Investigators may prefer a simpler approach and conduct interim analyses when all baskets at Stage 1 have finished accrual. This strategy will increase the trial duration. Temporarily, we assume some choice of baskets that make it past the interim analysis, $\bm{m}$. Moreover, assume the accrual rate is uniform through time. We focus here on the calculation of total trial duration, $d_{\bm{A}, \bm{m}}$. If all baskets or no baskets are pruned, the trial duration is the longest time it took a basket to finish accruing for the interim or end-of-trial sample size.

\begin{align*}
    \bm{m} = \{0\}^K &\implies d_{\bm{A}, \bm{m}} = \sup_{i\in \{1,...,K\}} \frac{N\cdot p_i \cdot t}{A_i}\\
    \bm{m} = \{1\}^K &\implies d_{\bm{A}, \bm{m}} = \sup_{i\in \{1,...,K\}} \frac{N\cdot p_i}{A_i}
\end{align*}

If some baskets are pruned, the calculation is more complicated. First, we calculate the time for each basket to reach the interim analysis. Consider the scenario where exactly one basket makes it past the interim analysis. We calculate the time needed to accrue to that basket to reach the initially planned final sample size (e.g. after re-allocating samples from pruned baskets). We calculate when the first basket will be pruned. If this duration is greater than the duration needed to accrue the final sample size, we add the ”waiting time” until the first basket is pruned to the current time the basket has already accumulated. Then update the sample size based on the basket being pruned and add the time it takes to accrue the new participants. We repeat the process for each basket that is pruned, either adding the time it takes to add the updated sample size or waiting time until the interim accruement finishes for the pruned basket. We repeat the process above for every choice of basket that makes it past the interim analysis to calculate the time it takes each basket to finish. The total trial duration, $d_{\bm{A}, \bm{m}}$, is the max duration from all the baskets that make it past the interim analysis.

\newpage
\bibliography{preprint_paper}

\end{document}